\documentstyle[multicol,prl,aps,epsfig]{revtex}
\begin{document}
\draft

\title{Distribution of fractal dimensions at the Anderson transition}
\author{D.~A.~Parshin$^{1,2}$ and H.~R.~Schober$^3$}
\address{$^1$Theoretical Physics Institute, University of Minnesota,
116 Church St. S.E., Minneapolis, Minnesota 55455}
\address{$^2$State Technical University, 195251 St.Petersburg, Russia}
\address{$^3$Institut f\"ur Festk\"orperforschung, Forschungszentrum
J\"ulich, D-52425, Germany}

\date{\today}
\maketitle
\begin{abstract}

We investigated numerically the distribution of participation numbers
in the 3d Anderson tight-binding model at the localization-delocalization
threshold. These numbers in {\em one} disordered system experience
strong level-to-level fluctuations in a wide energy range. The
fluctuations grow substantially with increasing size of the system.
We argue that the fluctuations
of the correlation dimension, $D_2$ of the wave functions are
the main reason for this. The
distribution of these correlation dimensions at the transition is
calculated. In the thermodynamic limit ($L\to \infty$) it does not
depend on the system size $L$. An interesting feature of this limiting
distribution is that it vanishes exactly at
$D_{\rm 2max}=1.83$, the highest possible value of the
correlation dimension at the Anderson threshold in this model. 

\end{abstract}
\pacs{63.50.+x,61.43.Hv,73.20.Jc,61.43.Fs}

\begin{multicols}{2}

The localization-delocalization Anderson
transition has posed for a long time a fascinating problem. 
In systems with short range interaction,
purely
diagonal disorder and unbroken time-reversal symmetry, without
spin-orbit interaction, it occurs for dimensions 
$d > 2$ \cite{LR}. In the thermodynamic limit (i.e.
infinite system size) the transition point separates the systems where
{\em all} wave functions are localized from the systems where 
some part of them is extended. Exactly at the transition 
one finds in the center of the band extended wave
functions. Due to the
proximity of the unavoidable localization
at one side of the transition they have a self-similar
fractal (actually multifractal) structure. This is a direct
consequence of quantum critical fluctuations at the transition point.

This multifractality of the wave functions at the localization
threshold is one of the most important features discovered 
\cite{Wegner,Aoki} since the
pioneering work of Anderson~\cite{A}.
This
fruitful idea has became widely recognized (see e.g.
Refs.~\cite{CP,J,FE}) and has considerably helped our understanding
of different phenomena in mesoscopic systems related to electron
localization. For instance the well known log-normal distribution of
the conductance in disordered metals~\cite{AKL} can be taken as a
finger-print of multifractal wave functions which 
survive in the weakly disordered state (so-called pre-localized
states~\cite{FE}).

Usually the multifractal (as well as fractal) structure of a wave
function manifests itself in the size dependence of the
participation number (PN)
\begin{equation}
{\cal N} = \left ( 
\int \left |\psi({\bf r})  \right |^4 d {\bf r}\right )^{-1}
\propto L^{D_2} ,
\label{eq:tr2}
\end{equation}  
where $L$ is the system size and $D_2 < d$ is the correlation
dimension of the wave function $\psi ({\bf r})$. For a localized state
$D_2=0$ and ${\cal N}$ do not depend on $L$. On the other hand $D_2=d$
for a delocalized wave function which extends uniformly 
over the sample. The inequality $0<D_2<d$ means
that a multifractal wave function is delocalized but, 
in the thermodynamic limit, nevertheless occupies
only an infinitesimal fraction of the sample.

Due to strong {\em level-to-level}
fluctuations\cite{YOY} it is
very hard  to verify the size dependence of ${\cal N}$ 
for a particular state in a computer experiment. At the transition
these fluctuations increase with increasing system size.
Therefore, one 
has to study the size dependence of the averages of ${\cal N}$ 
or, preferably,
the {\em distribution function}. The size dependence of 
the fluctuations of ${\cal N}$
is then converted  to the size dependence of
this distribution function. Choosing suitable size 
dependent variables one can 
collapse these distributions (for different system sizes) to one
universal curve. However, to do this systematically one needs to
understand the origin of these fluctuations. In our opinion
the main source of above mentioned giant fluctuations of PN
at the transition are the {\em fluctuations of the
fractal dimension} $D_2$ of the wave functions.

In a disordered system, exactly
at the transition~\cite{rem1}, critical wave functions
with very different degrees of delocalization (participation numbers)
should coexist
independent of their energy. This is a direct consequence of an
infinitesimal neighborhood of the localized behavior at one side of
the transition and delocalized behavior at another side. Accordingly
each delocalized wave function should have a different
$D_2$. If the
distribution function ${\cal P}(D_2)$
becomes size-independent in the thermodynamic limit our
hypothesis will be correct.

In the present paper we present numerical results for this
distribution (more exactly for the distribution of logarithms of
the PN) and show that it is indeed
universal, i.e. it does not depend on the system size in the
thermodynamic limit. This is somewhat reminiscent of the previous
idea of
Shapiro~\cite{BS,SC} about the existence of universal
distributions at the Anderson transition. This problem has 
recently been addressed
analytically in a couple of papers, but far from the
transition point.

In Ref.~\onlinecite{FM} the variance of the inverse
participation ratio (IPR) was calculated for disordered metallic
samples in leading order of small parameter $1/g^2$, where $g\gg
1$ is a dimensionless conductance of the system. It was suggested
that the relative value of the IPR-fluctuations
should be of order unity at the transition point and that their
distribution should be universal. A similar question was
raised  recently in Ref.~\onlinecite{PA} where the distribution of
the IPR was calculated for a
large but finite conductance $g$ of small metallic grains.

To investigate this problem at the transition we numerically solve
the standard Anderson tight-binding model with diagonal
disorder on a 3d simple cubic lattice with the Hamiltonian
\begin{equation}
{\cal H} = \sum\limits_i \varepsilon_i |i\left.\right >\left < \right.i| + 
\sum\limits_{i\ne j} t_{ij} |i\left. \right > \left < \right. j| .
\label{eq:ham1}
\end{equation}  
To model a disorder we distribute the site energies $\varepsilon_i$
uniformly in the interval $-W/2 < \varepsilon_i < W/2$.
For the off-diagonal elements we took $t_{ij}=1$ for nearest
neighbors and otherwise zero. The Anderson transition is in this model
at a critical value, $W_c = 16.5$ (see, e.g.
Ref.~\onlinecite{GS} and references therein).

\begin{figure}[htb]
\epsfig{figure=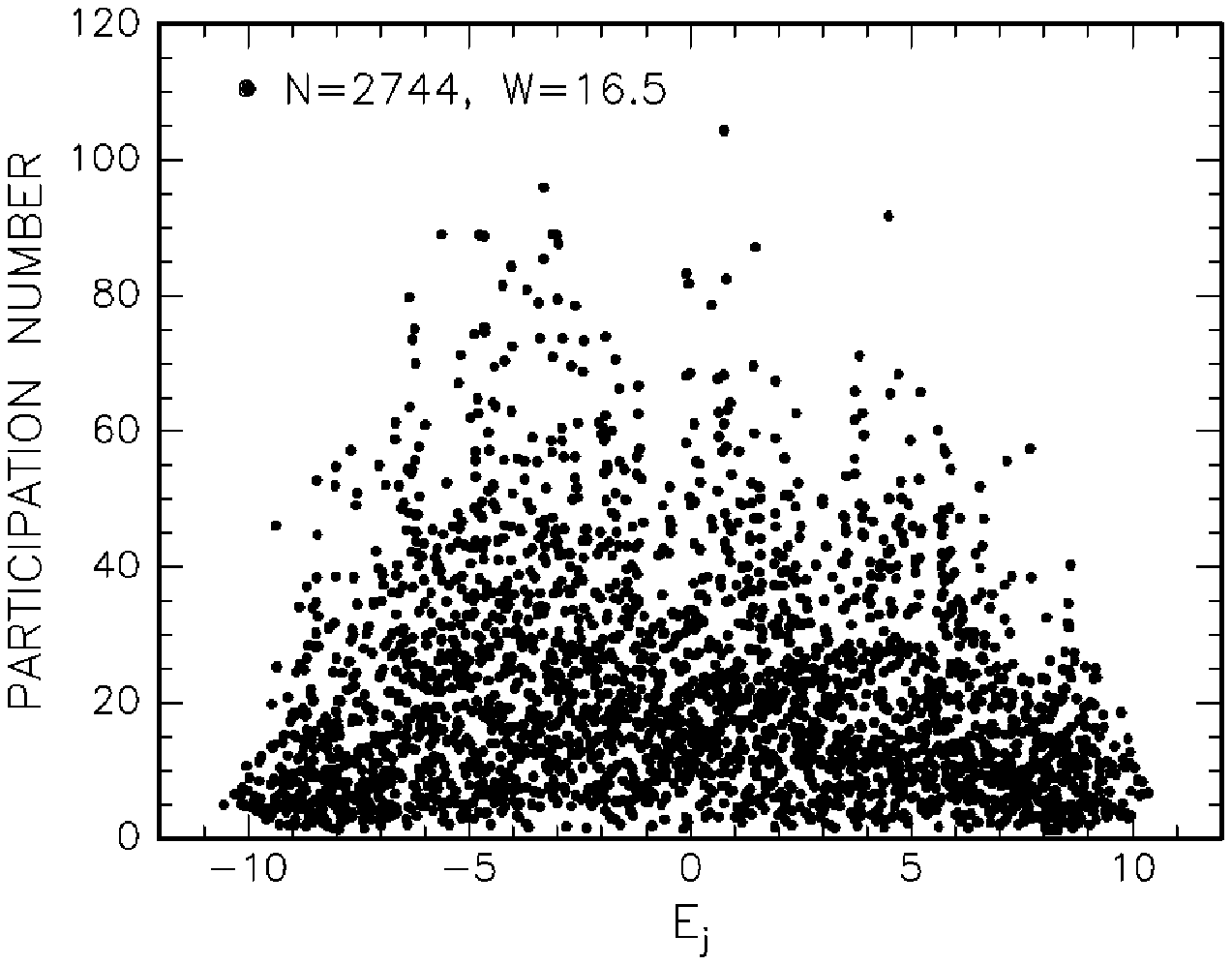,width=7.5cm,angle=0}
\caption{Participation numbers versus energy for a 3d system of
$14\times 14\times 14 = 2744$ sites at the Anderson transition.}
\label{fig:pnf1}
\end{figure}

Diagonalizing the matrix ${\cal H}_{ij}$ for a cubic sample with 
$N=L^{3}$ sites and open boundary conditions we find a set of $N$
orthonormal eigenvectors $e_s(j)$ 
and corresponding eigenvalues, $E_j$.
The participation number for a state $j$ is defined as usual by
\begin{equation}
{\cal N}_j=\left ( \sum\limits_{s=1}^N e_s^4(j) \right )^{-1}.
\label{eq:pr}
\end{equation}

Fig.~\ref{fig:pnf1} shows the participation numbers ${\cal N}_j$
versus energy $E_j$ for a 3d system with $N=2744$ sites. 
At the transition there is a strong level-to-level fluctuation of the PN.
In the whole energy range states are found 
which are  very close in energy but whose ${\cal N}_j$
differ by up to two orders of magnitude. This means that the energy of a
particular state is not indicative of the spatial behavior of
the wave function. 
In a next step we, therefore, calculate the distribution function
of the PN in an energy band around zero
energy (where the Anderson transition takes place). We take the width
of this band equal $10$ i.e. we include all states with
$|E_j|<5$. The distribution remains practically the
same if we take a much narrower strip, $|E_j|<1$ but the 
computation time increases significantly. 

\begin{figure}[htb]
\epsfig{figure=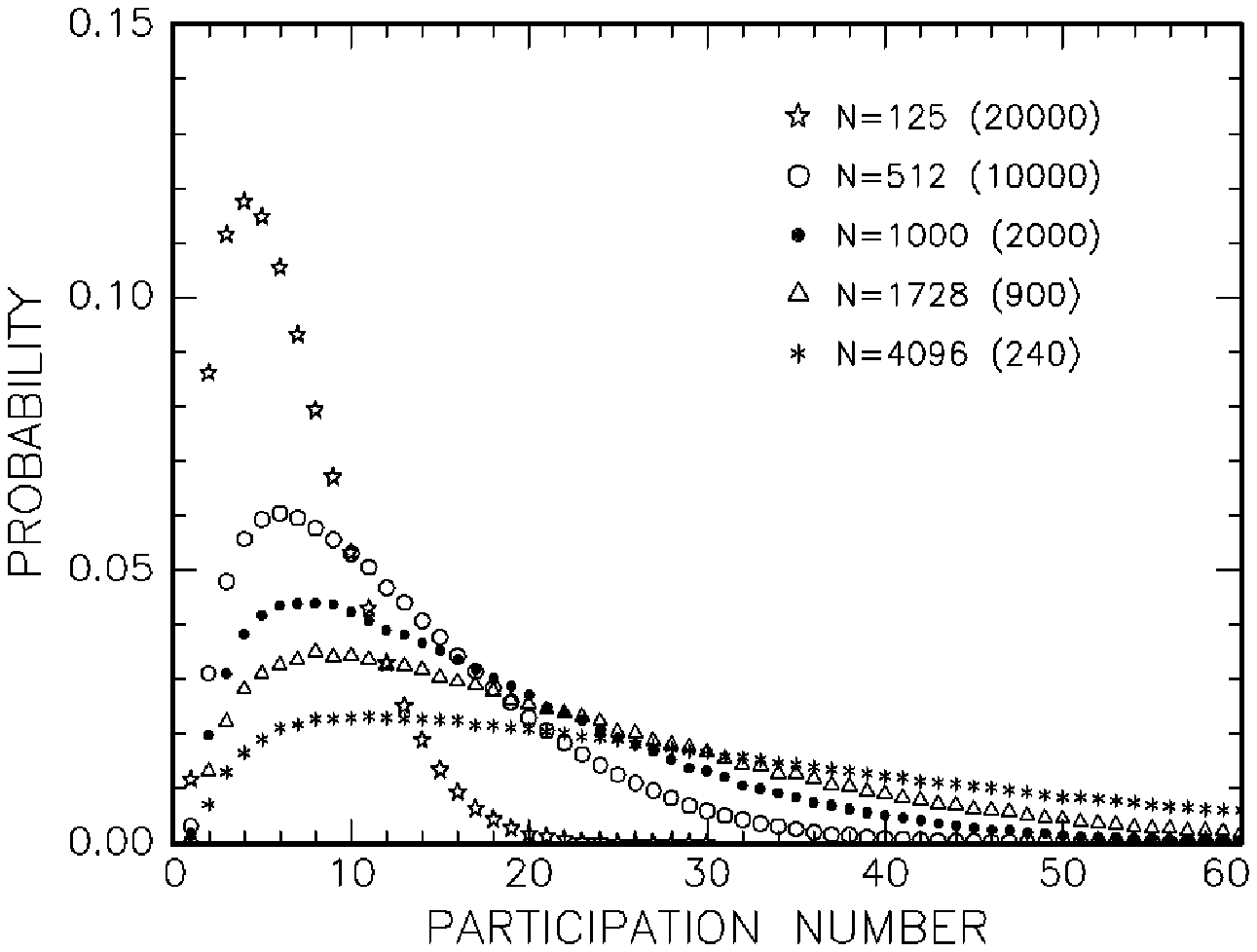,width=7.5cm,angle=0}
\caption{Distribution functions of participation numbers at the
Anderson transition for different system sizes. In the brackets the
numbers of realizations used in the averaging are
shown} 
\label{fig:tre1}
\end{figure}

The normalized distribution functions of the participation numbers for
different system sizes are shown in Figure~\ref{fig:tre1}. As expected
the distributions are strongly size dependent. Increasing the
system size the position of the maximum shifts to higher values,
approximately as $\propto N^{0.3}$, and the
amplitude of the maximum decreases as $\propto N^{-0.46}$.

\begin{figure}[htb]
\epsfig{figure=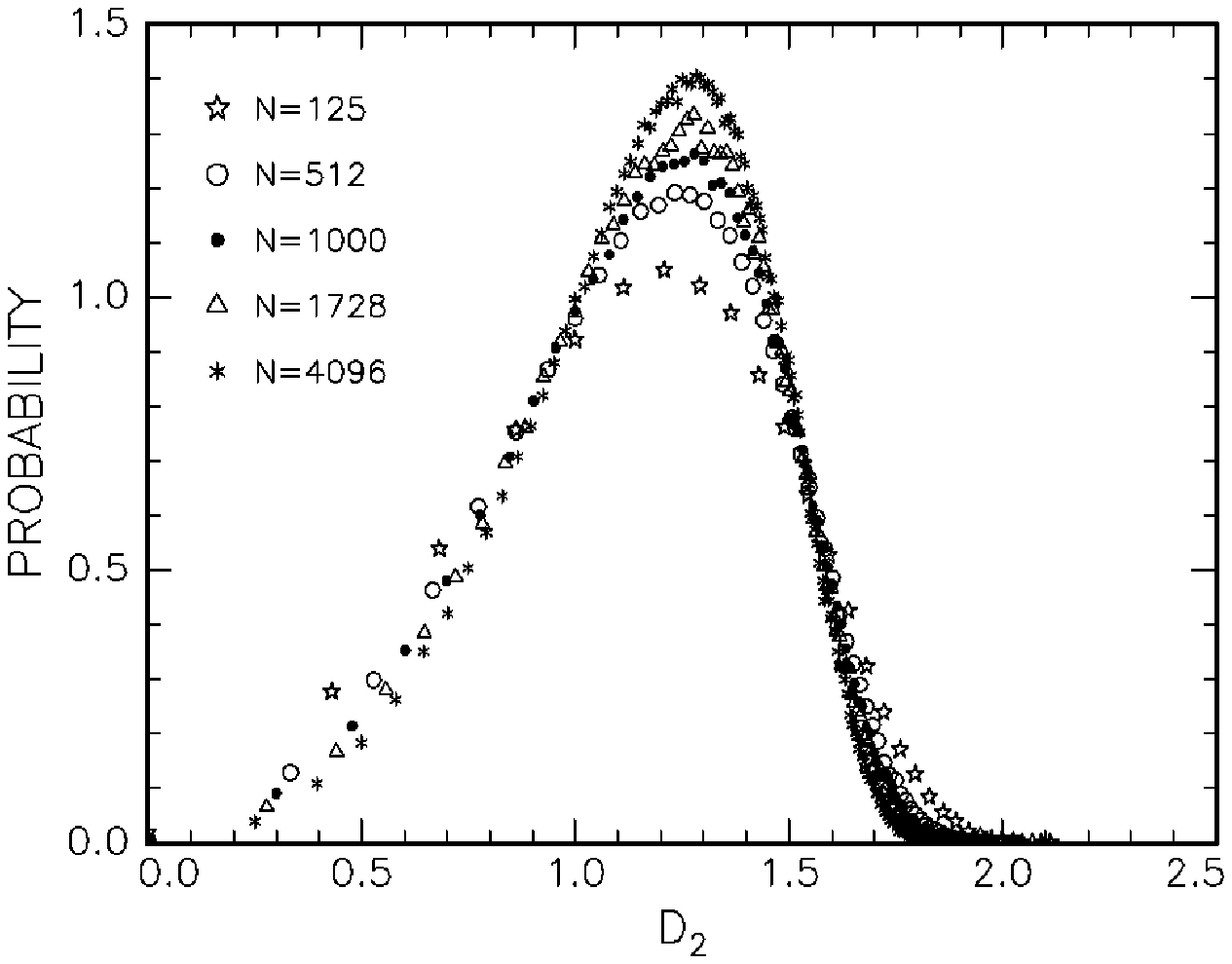,width=7.5cm,angle=0}
\caption{Distribution functions of the correlation dimension ${\cal
P}_N(D_2)$ at the Anderson transition for different system sizes.}
\label{fig:lpn1}
\end{figure}

Let us suppose now that the fluctuation of the
PN at the transition is due to fluctuations of 
$D_2$ of the wave functions.  According
to Eq.~(\ref{eq:tr2}), without loss of generality, we can take this
relation in the form ${\cal N}\propto N^{D_2/3}$. 
Then if $F_N({\cal N})$ is
the normalized distribution function of PN for an
ensemble of systems with $N$ sites and different disorder, we
can extract the distribution function of the
correlation dimensions, $D_2$, in this ensemble
\begin{equation}
{\cal P}_N(D_2) = (1/3) F_N\left ( N^{D_2/3}\right ) N^{D_2/3}\ln N .
\label{eq:cddf1}
\end{equation}  
Fig.~\ref{fig:lpn1} shows that this distribution of correlation dimensions
is much less sensitive to the system size than the one of the PN.
Moreover, with increasing $N$ it obviously approaches
some size independent function, ${\cal P}_{\infty}(D_2)$, i.e.
in the thermodynamic limit a universal distribution function of
correlation dimensions does exist at the Anderson transition.
For a given model of disorder it should be the
same for different realizations. In other words we believe that the
distribution function ${\cal P}_{\infty}(D_2)$ is a {\em
self-averaged} quantity~\cite{rem2} and can be obtained by  analyzing
one sufficiently large system.

Fig.~\ref{fig:lpn1} shows an additional interesting feature of
this distribution: with
increasing argument the function ${\cal P}_N(D_2)$ 
rapidly decays to zero. The drop gets steeper with increasing size
$N$. 
This is seen more clearly in a logarithmic plot of
${\cal P}_N(D_2)$. 
For systems with $N\geq 512$, ${\cal P}_N(D_2)$
drops four orders of magnitude in the small
interval $1.5<D_2<2$. We think, therefore, that in the
thermodynamic limit there is a point, $D_{\rm 2max}$ where the
function ${\cal P}_{\infty }(D_2)$ approaches {\em exactly zero}. In
such a case this would be the highest possible value of the correlation
dimension $D_2$ at the Anderson transition. In the following 
we present additional
evidence in  support of  this idea.

\begin{figure}[htb]
\epsfig{figure=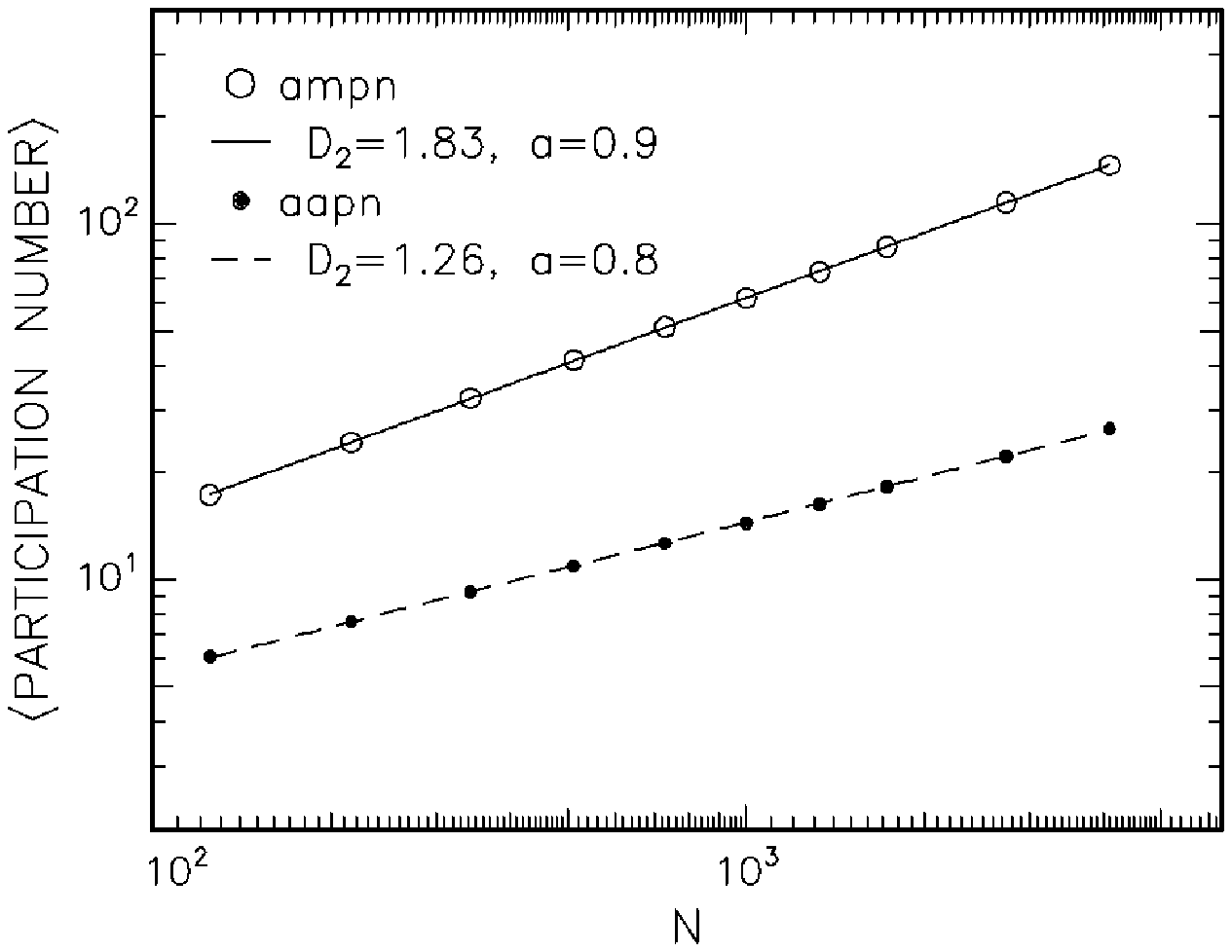,width=7.5cm,angle=0}
\caption{Averaged maximum participation number ({\em ampn}) and averaged
average participation number ({\em aapn})  at the
transition versus system size. 
Lines are the best least squares fit with 
$\left<{\cal N}\right> = aN^{D_2/3}$, where $\left<{\cal N}\right>$
is "ampn" or "aapn", respectively.} 
\label{fig:aapn}
\end{figure}

For that purpose we calculate the size dependence of the average PN and
of the PN of the
the {\em most extended state} in
the system, i.e. for a given realization of disorder 
the state with maximal participation
number, ${\cal N}_{\rm max}$.
The ensemble averages of both quantities (denoted by
{\em aapn} and {\em ampn}, respectively) 
against system size $N$ are plotted in Fig.~\ref{fig:aapn}.
First, both quantities scale with $N$
as a power law, $aN^{D_2/3}$. Secondly
the value $D_2=1.26$ for the average participation
number ({\em aapn}) is very close to the position of the maximum in the
distribution function of correlation dimensions shown in 
Figure~\ref{fig:lpn1}. It is surprising that averaging over all
states does not affect the power law behavior given by
Eq.~(\ref{eq:tr2}) for a single state. 
For the ensemble average of the 
maximum participation number, $\left<{\cal N}_{\rm max}\right>$
we find  $D_2=1.83$.
This is the correlation dimension of the most
extended state in the system at the transition. To answer the
question whether it really is the maximal correlation dimension, $D_{\rm 2max}$,
we should study the fluctuations of this quantity.
A {\em maximal} correlation dimension
$D_{\rm 2max}$ does exist if the fluctuation of this quantity goes to zero
with $N\to\infty$.  

\begin{figure}[htb]
\epsfig{figure=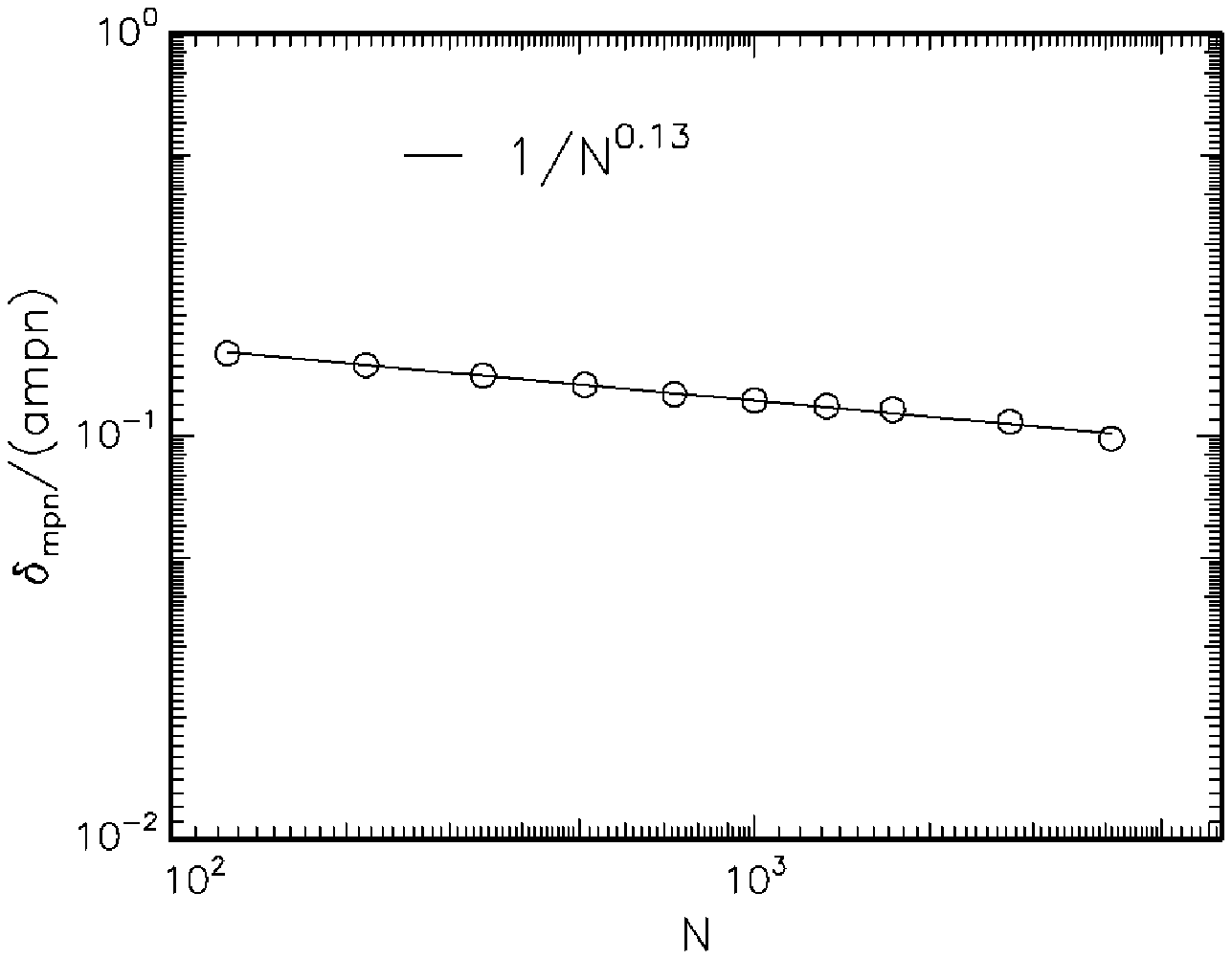,width=7.5cm,angle=0}
\caption{Relative root mean square deviation of the maximum
participation number against system size. The solid fitting line
corresponds to a power law behavior, $\propto 1/N^{0.13}$.}
\label{fig:dmpn}
\end{figure}

Fig.~\ref{fig:dmpn} shows the relative fluctuations of the maximum
participation number versus system size. Firstly
the fluctuations are rather small and {\em decrease} 
slowly with increasing
$N$. According to Eq.~(\ref{eq:tr2}) we can relate them to
the fluctuations of the correlation dimension $\delta D_{\rm 2max}$ 
\begin{equation}
\delta_{\rm mpn}\equiv 
\delta {\cal N}_{\rm max}/\left<{\cal N}_{\rm max}\right> =
(\delta D_{\rm 2max}/3) \ln N .
\label{eq:rfn1}
\end{equation}
We conclude from the above that, in the thermodynamic
limit ($N\to \infty$), the fluctuation of the correlation dimension for
the most extended state 
$\delta D_{\rm 2max}\to 0$ at least not slower than $1/\ln N$.
Therefore $D_{\rm 2max}=1.83$ is indeed the highest possible value of
the wave function correlation dimension at the Anderson transition.
The distribution function ${\cal P}_{\infty}({\cal D}_2)$ 
should approach exactly zero at this value. 

There are numerous previous 
calculations of the correlation dimension $D_2$ 
at the 3d Anderson transition.
From the scaling with system size of
the density-density correlation of the wave functions, the correlation
dimension was estimated to be $D_2=1.7\pm 0.3$ (for $W=16$)\cite{SE}.
%
From the size dependence of the participation
number averaged, both  over different wave functions in a small energy
interval $0.25$ around zero and over disorder (with a Gaussian
distribution of site energies), the correlation dimension $D_2=1.6\pm
0.1$ was estimated\cite{S1}. 
%
In Ref.~\onlinecite{BMP} 
from the spectral compressibility, $\chi$ of the levels, using
$\chi = (1-D_2/d)/2$ derived in Ref.~\onlinecite{CKL} the correlation 
dimension was estimated as $D_2=1.4\pm 0.2$.
From the time decay of the temporal autocorrelation
function, $C(t)\propto t^{-D_2/3}$, the correlation dimension, $D_2=1.5\pm 0.2$
was obtained\cite{OK}.
Using box-counting procedures $D_2=1.7\pm 0.2$ \cite{BHS}, 
$D_2=1.52\pm 0.11$ \cite{TT}, and $D_2=1.68$ \cite{SG} was calculated.
Again using box counting and averaging the results over wave functions 
in a small energy
interval, $\Delta E = 0.01$, around zero and over five different
realizations of disorder $D_2=1.46$ was found \cite{GS}.
Again using box-counting techniques in
Ref.~\onlinecite{Ev}, $D_2=1.33\pm 0.02$ was obtained.

This shows clearly the quite large uncertainty, in the literature, 
in the existing
values of $D_2$ at the transition. The reason is
obviously that one deals with a distribution of
this quantity. Among different characteristics of this distribution
one can consider for example the position of the maximum, at about
$1.3$, the average value, $\overline{D_2}=1.26$, and the correlation
dimension of the most extended state at the transition, $D_{\rm
2max}=1.83$.

Similarly the distribution function of the information
dimension $D_1$ as well as of the other generalized dimensions $D_q$ at the
transition can be obtained. The results of this analysis will be
published elsewhere. In the thermodynamic limit all these
distribution functions are expected to be universal and
self-averaged quantities. Using the 
Legendre transformation the multifractal spectrum $f_j(\alpha)$ for
{\em each} wave function with energy $E_j$ can be calculated (compare
Ref~\cite{GS2}). In other words critical wave functions at the
transition should be characterized by a whole distribution of
$f(\alpha)$'s which is already a {\em functional}. In
particular, the positions of the maxima, $\alpha_0$'s
of these functions should be distributed.

In conclusion, we investigated numerically the distribution of the
correlation fractal dimension $D_2$ for the critical wave functions
at 3d Anderson transition. This distribution appears to be
universal i.e. it no longer depends on the system size in the thermodynamic
limit.  Extrapolating these results to other fractal dimensions we
conclude that each critical wave function should possess its own
(infinite) set of generalized fractal dimensions, $D_d$ at the
transition. And one should rather speak about the distribution
functional of these functions.

The authors are very grateful to B.I. Shklovskii for fruitful
discussions. One of us (D.A.P.)
gratefully acknowledges the financial support and hospitality of
the University of Minnesota where part of this work was done.  


\end{multicols}

\begin{references}
\bibitem{rem1} In fact it is sufficient that the correlation length $L_c$
was much bigger than the sample size $L$.
\bibitem{rem2} To calculate the distribution functions
${\cal P}_N(D_2)$ for the rather small sizes shown in
Fig.~\ref{fig:pnf1} we had, of course, to average over many different
realizations of disorder.
\end{references}
\end{document}